# Anisotropic Elastic Modulus, High Poisson's ratio and Negative Thermal Expansion of Graphynes and Graphdiynes

*Sergio A. Hernandez, Alexandre F. Fonseca* [*]

Applied Physics Department, Institute of Physics "Gleb Wataghin", University of Campinas - UNICAMP, 13083-859, Campinas, São Paulo, Brazil.

**Corresponding Author**

[*] Phone: +55 19 3521-5364. Email: Alexandre F. Fonseca – afonseca@ifi.unicamp.br

**ABSTRACT:** Graphyne (GY) and graphdiyne (GDY) are two-dimensional one-atom-thick carbon allotropes highly considered to substitute graphene in electronic applications because of the prediction of non null band-gap. There are seven types of GY structures not yet fully investigated in literature. In this work, by means of classical molecular dynamics simulations, the Young's modulus, Poisson's ratio and linear thermal expansion coefficient (TEC) of all originally proposed seven types of GYs and corresponding GDYs are calculated. The dependence of these properties with the density of the structure is investigated for the first time. Quadratic increasing of the TEC of GY and GDY structures with density was found. The elastic modulus of GYs and GDYs were shown to be more sensitive to their density than general porous materials. In particular, non-symmetric structures are much softer along the armchair direction than along zigzag direction, implying that the elasticity along armchair direction of GY and GDY structures are similar to that of porous gels materials. Values larger than unity were found for the Poisson's ratio of some non-symmetric GYs and GDYs. A simple honeycomb mechanical model is shown to capture the observed values of Poisson's ratio of GYs and GDYs.

## I. INTRODUCTION

Graphyne (GY) and graphdiyne (GDY) are two-dimensional one-atom-thick carbon allotropes whose structures can be imagined as simply formed by coherently replacing certain carbon-carbon bonds in a graphene hexagonal network by acetylenic (–C≡C–) and diacetylenic (–C≡C–C≡C–) linkages, respectively [1,2]. Although proposed in the 80's by Baughman *et al.* [3], GYs have got increasing interest in the last few years mainly because of the prediction of non-null band-gap for some types of GYs [1-7]. This indicates the possibility to use them in electronic nanodevices instead of graphene [7]. Generally speaking, GYs can be good electrical conductors or semiconductors [3-10], are predicted to present higher conductivity than graphene [11,12], have smaller in-plane stiffness than graphene, what depends on the number of acetylenic or diacetylenic linkages [13-26], and have much lower thermal conductivity than graphene [9,27-30]. Good electrical and poor thermal conductivity form a good combination for possible applications of GYs in thermoelectrics [27-28]. Besides, the presence of acetylenic and diacetylenic linkages introduces two-dimensional porosity to the GY and GDY structures. Combinations of all these interesting properties have inspired ideas on possible applications of GYs and GDYs such as on manufacturing nanotransistors, anodes in batteries, desalination, hydrogen storage, among others [1,2,31-35].

GYs have not been yet synthesized, but subunits of GY and GDY structures have been produced as reported in the literature [36-39]. The synthesis of GDYs has been claimed for a nanoscale film [40], tubes [41], and ordered stripe arrays [42]. Recently, the production of very thin stacks (24 nm thick) of multilayered GDY nanosheets has been reported [43]. It is worth to mention the work of Han *et al* [44]. which used first-principles methods to derive a chemical potential phase diagram of GYs and graphene in order to suggest the best experimental conditions for the synthesis of GYs.

Graphyne-based structures such as nanoribbons [45,46], nanotubes [47-52], AB stacking of α-graphynes (γ-graphynes) or the so-called "AB-α-graphityne" ("AB-γ-graphityne") [53], as well as functionalized GYs [54-57] have been also subject to study, mostly using first-principles methods.

In the literature, the most studied GY structures are the ones with prefixes called "γ-", "6,6,12-", "β-", and "α-" graphynes. Most of the papers about these structures performed their studies on individual GYs or GDYs, or families of the same type of graphynes having *one*, *two*, *three* or *more* adjacent acetylenic linkages (graphyne, graph*di*yne, graph*tri*yne, or graph"*more*"yne, respectively). However, although more recent studies have presented comparisons of some physical properties of more than one type of GY (see, for example, the Refs. [7,10,16,25]), here we are concerned with a more comprehensive and comparative study of the Young's modulus (or in-plane stiffness in units of force per unit length), Poisson's ratio and linear thermal expansion coefficient (TEC) of all seven types of GYs, as originally proposed by Baughman *et al*. [3], and the corresponding GDYs.

Figure **1** shows the GY structures with the numbering notation defined by Ivanovskii [1] that will be adopt here for simplicity. The corresponding GDYs can be formed by changing the acetylenic to diacetylenic linkages and will have the same numbering notation.

Regarding the mechanical properties of graphynes, Chen *et al*. [11], Cranford and Buehler [13], and Zhang *et al.* [16] were the first to calculate the elastic modulus and/or fracture properties along armchair and zigzag directions of γ-, β- and α-graphynes (our GY1, GY4 and GY7 structures). Additionally, Chen *et al*. [11] and Zhang *et al*. [16] observed a strong anisotropy in the mechanical properties of the 6,6,12-graphyne (our GY2 structure). This result agrees with the observation that only 2D structures with square or hexagonal symmetries present symmetric elastic constants along armchair and zigzag directions [58]. This anisotropy was also observed in the electrical properties of 6,6,12-graphyne (or GY2) [7,10]. In speaking of anisotropy, Cranford and Buhler [13] reported about 18% of difference between armchair and zigzag Young's modulus of γ-graphyne. As γ-graphyne has hexagonal symmetry, this difference should not be so high.

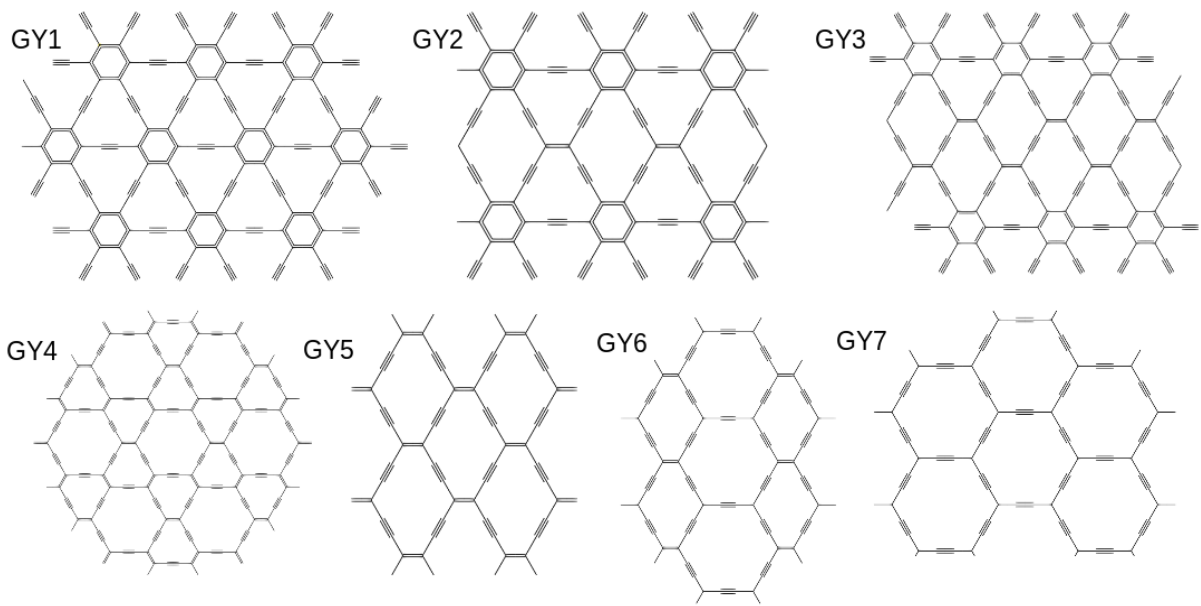

**Figure 1**: The seven GY structures proposed by Baughman *et al.* [3] Horizontal (*x*) and vertical (*y*) axes are called "armchair" and "zigzag", respectively. The numbering notation and sequence are the same as defined by Ivanovskii [1]. In the literature, GY1, GY2, GY4 and GY7 are called "γ-", "6,6,12-", "β-", and "α-" graphynes, respectively. Along the text, we will make use of both notations when comparing our results to those from literature. Corresponding GDYs from GDY1 to GDY7 are similar but with diacetylenic linkages instead of acetylenic ones.

Here, we investigate the anisotropy in the mechanical properties of all seven types of GYs and corresponding GDYs. As they are more porous than graphene, we also investigate their elastic modulus-density relationship. Non-symmetric GY and GDY structures will be shown to present an interesting dependence of their mechanical behavior on the direction along which the stress is applied.

Another interesting property of GYs, not fully explored yet, is the thermal expansion. The thermal expansion relates the variation of the dimensions of an object to the variation of temperature. It is known that graphene presents negative in-plane thermal expansion coefficient [59-67]. This thermal contraction has been attributed to large negative Grüneisen parameters related to out-of-plane acoustic modes that are predominantly excited at low temperatures [59]. Acoustic out-of-plane transversal modes are easily excited because of the planarity of graphene [65]. As graphynes are also two-dimensional structures, it is expected that they present negative TEC. To our knowledge, there are only three works reporting the calculation of the linear thermal expansion coefficient of graphynes [10,17,68] and except for the work of Reference [17] the other two [10,68] reported values of graphyne TEC up to one order of magnitude larger, in modulus, than that of graphene, at temperatures below or close to room temperature. All of these works used first-principles methods to calculate the TEC of graphynes. Here, we present the calculations of the linear TEC of all seven types of GYs and corresponding GDYs using classical molecular dynamics (MD) simulations. We show that the thermal contraction of GYs and GDYs are one order of magnitude larger than that of graphene.

We also calculated the Poisson's ratio of GYs and GDYs. Literature reported calculations of Poisson's ratio of α-graphyne [21,25,26], β-graphyne [25], γ-graphyne [18,20,24,25,69], α-graphdiyne [21,26], γ-graphdiyne [14,20,58] and the graphite-like structures AB-α-graphityne and AB-γ-graphityne [53], most of them from first-principles methods. Couto and Silvestre [24] reported a difference of about 7% between the calculated Poisson's ratio along two different directions of γ-graphyne. Here, by employing a methodology that does not assume any symmetry of the structures (as explained ahead), we verify these results, reveal the isotropy or anisotropy in the elastic properties of not yet studied GYs and GDYs, and show that some structures also possess values of Poisson's ratio above unity. A honeycomb mechanical model is considered to interpret these results for the Poisson' ratio of GYs and GDYs. Table **I** presents our results together with the existing literature values of Young's modulus and Poisson's ratio of all GY and GDY structures studied here.

## II. THEORY AND COMPUTATIONAL METHODS

**Linear Thermal Expansion Coefficient (TEC)**. The linear TEC, α, of a material is defined by:

$$\alpha_K(T) = \frac{1}{L}\left(\frac{dL}{dT}\right)_P , \qquad (1)$$

where $L = L(T)$ is the equilibrium length of the material at the absolute temperature $T$, under constant pressure $P$. Here, our calculations are made at zero pressure and we will omit this dependence from now on.

In order to calculate the TEC of the GYs and GDYs, the equilibrium lengths of the structure along $x$ (armchair) and $y$ (zigzag) directions as functions of the temperature $T$, $L_x(T)$ and $L_y(T)$ respectively (see figure **1** for the definition of "$x$" and "$y$" directions), must be obtained. Then, we use equation (1) to obtain the corresponding $\alpha(T)$. The equilibrium values of $L_x(T)$ and $L_y(T)$ will be obtained using classical molecular dynamics (MD) simulations as described below.

**Computational methods for calculation of TEC.** The protocols for the calculations of the linear TEC of GYs and GDYs consist of performing long MD simulations of each structure at zero pressure and at a given value of temperature from a list of previously chosen values of temperature in the range between 50 K and 1000 K. From every MD simulation performed at a given temperature, one equilibrium value of the lengths $L_x$ and $L_y$ of the structure is obtained. The list of equilibrium values of $L_x$ and $L_y$ corresponding to the list of previously chosen values of temperature is further used to obtain $L_x(T)$ and $L_y(T)$ through a linear regression function in $T$. This protocol to obtain the TEC of GYs is the same one used to obtain the linear TEC of graphene nanomeshes [66] and large bilayer graphene structures [67].

All simulations were performed using the LAMMPS package [70], Langevin [71] thermostat and Berendsen [72] barostat. Langevin (Berendsen) damping factor was set to 1 ps (10 ps). As the TEC of graphene depends on the structure size [66,67], samples of six different sizes of each GY and GDY structure were generated to test the convergence of the calculations of the linear TEC with system size. The results of this test are shown in figures **S1** and **S2**, respectively, of Supplementary Material [73]. The system sizes of about 600 Å were chosen for all GYs and GDYs so as to make meaningful comparisons. The simulations were performed with periodic boundary conditions (PBC) along $x$ and $y$ directions, and the box sizes were considered as the structure sizes. The integration algorithm used was verlet with a timestep of 0.5 fs. Each MD simulation at a given value of temperature was performed by the total of 4 x $10^6$ timesteps (or 2 ns) to ensure both thermal and pressure equilibrations, with pressure set to zero and the box sizes allowed to relax. During every MD simulation, instantaneous values of the structure along $x$ and $y$ directions, $l_x$ and $l_y$, respectively, were exported at each 200 timesteps (or 100 fs). The equilibrium values of $L_x$ and $L_y$ at a given temperature were, then, obtained from the average of $l_x$ and $l_y$, respectively, taken from the last 2 x $10^6$ timesteps (or 1 ns) of simulation. The square root of the variance of this average was taken as uncertainty of $L_x$ and $L_y$.

The AIREBO potential [74,75] was employed to simulate the carbon-carbon interactions. It has been used to study mechanical and thermal properties of GYs and GDYs [15,16,22,45,76]. Yang and Xu [15], for example, have compared AIREBO results for the equilibrium bond distances of different sp and $sp^2$ carbons in GY structures with those from DFT and other classical MD potentials, showing excellent agreement between them.

**Theory: Young's modulus and Poisson's ratio**. The in-plane mechanical properties of GYs and GDYs will be obtained from the calculation of the elastic constants $C_{11}$, $C_{12}$ and $C_{22}$ of the structures as described by Andrew *et al.* [58] Basically, the elastic energy per unit area of the structure is given by:

$$U(\varepsilon) = \frac{1}{2}C_{11}\varepsilon_{xx}^2 + \frac{1}{2}C_{22}\varepsilon_{yy}^2 + C_{12}\varepsilon_{xx}\varepsilon_{yy}, \qquad (2)$$

where $\varepsilon_{xx}$ and $\varepsilon_{yy}$ are strains along $x$ and $y$ directions, respectively. Once $C_{11}$, $C_{12}$ and $C_{22}$ of the structures are obtained, the Young's modulus and Poisson's ratio of the structures are given by:

$$E_x = \frac{C_{11}C_{22}-C_{12}^2}{C_{22}}, \quad E_y = \frac{C_{11}C_{22}-C_{12}^2}{C_{11}}, \qquad (3)$$

$$\upsilon_{xy} = \frac{C_{12}}{C_{22}}, \quad \upsilon_{yx} = \frac{C_{12}}{C_{11}}, \qquad (4)$$

where $E_x$ ($E_y$) is the Young's modulus of the structures along $x$ ($y$) direction, and $\upsilon_{xy}$ ($\upsilon_{yx}$) is the Poisson's ratio of the structures when longitudinal strains are applied along $x$ ($y$) direction and transversal strains are observed along $y$ ($x$) direction, respectively. We are not going to previously assume that the structures are or not symmetric so as to verify the assertion that Young's modulus and Poisson's ratio of symmetric structures are equal along *x* and *y* [58]. In other words, we are going to calculate $C_{11}$ and $C_{22}$ separately not previously assuming they are equal as some studies do. This is also a way to test the validity of our methods.

In order to interpret the results, we derived expressions for the Poisson's ratio of a non-symmetric honeycomb structure as shown by figure **2**, inspired by one model proposed by Grima *et al.* [77]. Grima *et al.* have derived expressions for the Young's modulus and Poisson's ratio of a honeycomb structure due to hinging or stretching of its ribs. A combination of both types of deformations lead to the following expressions for the Poisson's ratio along two directions, $v_{xy}$ and $v_{yx}$:

$$v_{xy} = \frac{2(h/l+1/2)}{3K_S+K_H(8h/l+1)}(K_S - K_H) \text{ and } v_{yx} = \frac{3/2}{(h/l+1/2)(K_S+3K_H)}(K_S - K_H) \quad , \tag{5}$$

where *l* and *h* are the sizes of the ribs, the angle *θ* (see figure **2**) between the ribs was considered to be 120° in the unstressed state, and $K_S$ and $K_H$ are the elastic force constants corresponding to the stretching and hinging of the ribs, respectively. It is easy to show that if *l* = *h*,

$$v_{xy} = v_{yx} = \frac{K_S - K_H}{K_S + 3K_H} \quad , \tag{6}$$

consistent with what is expected for structures with hexagonal symmetry [58]. Equation (6) also predicts that the Poisson's ratio of symmetric structures cannot be larger than 1.

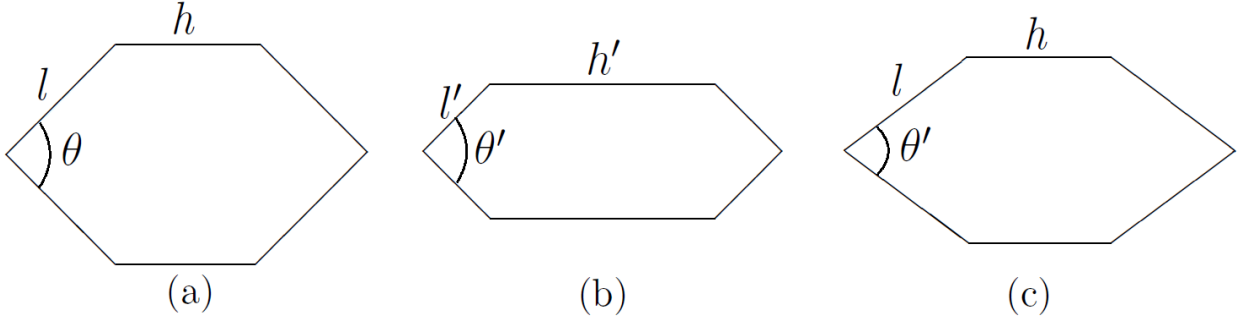

**Figure 2**: (a) Honeycomb cell model formed by two ribs of sizes *l* and *h*, and an opening angle *θ*. (b) Deformed cell by stretching/compression of the ribs. (c) Deformed cell by hinging. Deformations are exaggerated for clarity.

**Computational methods for elastic properties.** In order to obtain $C_{11}$, $C_{12}$ and $C_{22}$ for each structure, we will perform three series of simulations: two *uniaxial* tensile tests and one *biaxial* tensile test. The uniaxial tensile test consists of fixing one dimension, for example, the length of the structure along *y* direction, thus making $\varepsilon_{yy} = 0$, and then obtaining the energy of the structure as a function of strain along the other direction, through equation (2): $U^{\text{uniaxial}} \equiv U(\varepsilon = \varepsilon_{xx}, \varepsilon_{yy} = 0) = 0.5 C_{11}\varepsilon^2$. This will allow to obtain $C_{11}$. Conversely, by fixing the length of the structure along *x* direction, thus making $\varepsilon_{xx} = 0$, we can obtain $C_{22}$. The biaxial test consists of applying the same amount of strain along *x* and *y* directions at the same time, i. e., by making $\varepsilon_{xx} = \varepsilon_{yy} \equiv \varepsilon$, so that $C_{12}$ can be obtained, provided the $C_{11}$ and $C_{22}$ were already calculated. In this case, equation (2) becomes: $U^{\text{biaxial}} \equiv U(\varepsilon = \varepsilon_{xx} = \varepsilon_{yy}) = M\varepsilon^2$, and $C_{12} = M - 0.5(C_{11} + C_{22})$.

The tensile tests were made through energy minimizations of the structure under periodic boundary conditions and with applied strains consistent with the type of tensile test. Each energy minimization was performed by the conjugate-gradient algorithm implemented in LAMMPS, with energy and force tolerances set to 0.0 and $10^{-8}$, respectively, in order to ensure the convergence of the structure to the minimized energy one.

**III. RESULTS and DISCUSSION**

**Thermal Expansion Coefficient (TEC) of GYs and GDYs**

The study of the thermal expansion of nanostructures is important to quantify the amplitude of thermal effects on the shape and size of the structure. This property has been not well explored in GYs and there is not any estimation of the TEC of GDYs as well as a comparison amongst them.

Here, the TEC of all seven GYs and corresponding GDYs were calculated following the protocols described in the previous Section. The values of the equilibrium lengths along $x$ and $y$ directions of all seven GY and GDY structures were obtained from the MD simulations as function of temperature (Figures **S3** and **S4** of Supplementary Material [73]). The choice of sizes about 600 Å happened after testing the convergence of the TEC as function of the structure size (see Figures **S1** and **S2** of Supplementary Material [73]). Using Eq. (1), the TEC as function of temperature of all structures were calculated (Figures **S5** and **S6** of Supplementary Material [73]). We can see that the values of the TEC of the symmetric structures (GY1, GDY1, GY4, GDY4, GY7 and GDY7), as functions of temperature, along $x$ and $y$ directions are close to each other as expected. We calculated the same results for a graphene sample of similar size in order to have a reference for comparison (left bottom corner of Figures **S3** and **S5** of Supplementary Material [73]).

Figure **3** presents the results for the average TEC at 300 K of all GYs and GDYs as function of the density, $\rho$, of the structure relative to that of graphene: $\rho = \rho_{(GY\ or\ GDY)}/\rho_{GRAPHENE}$. The first detail observed in Figure **3** is that the TEC of all GYs and GDYs are, in modulus, about one order of magnitude larger than that of graphene. This result reasonable agrees with those obtained from first principles methods [10,17,68]. Perkgöz and Sevik [10] and Kim *et al*. [68] reported large values of TEC of graphynes in the same order of magnitude of our results, but for temperatures below room temperature. This difference might be a consequence of the classical method employed in the calculations that does not capture quantum mechanics effects. Besides, the first principle methods used to predict the TEC of graphene [59], resulted in a value about half the one measured experimentally [60]. On the other hand, our results are consistent with the existence of rigid unit modes (RUMs) that were shown to strongly contribute to the thermal contraction of the structures [68].

A second result from Figure **3** is an approximately quadratic dependence of the TEC of the structures with the relative density. This result is consistent with the above mentioned contribution of the RUMS of GYs and GDYs to their TEC. According to Kim *et al*. [68], the stretching modes of carbon-carbon triple bonds are relatively protected against scattering by large energy gaps. Therefore, thermal fluctuations will excite more intensely the RUMs. Smaller the density, larger the room for increasing amplitudes of these RUMs. It will be interesting to see up to which number of acetylenic linkages the TEC of the corresponding graph"*more*"ynes will continue increase in modulus. This will be subject of future work.

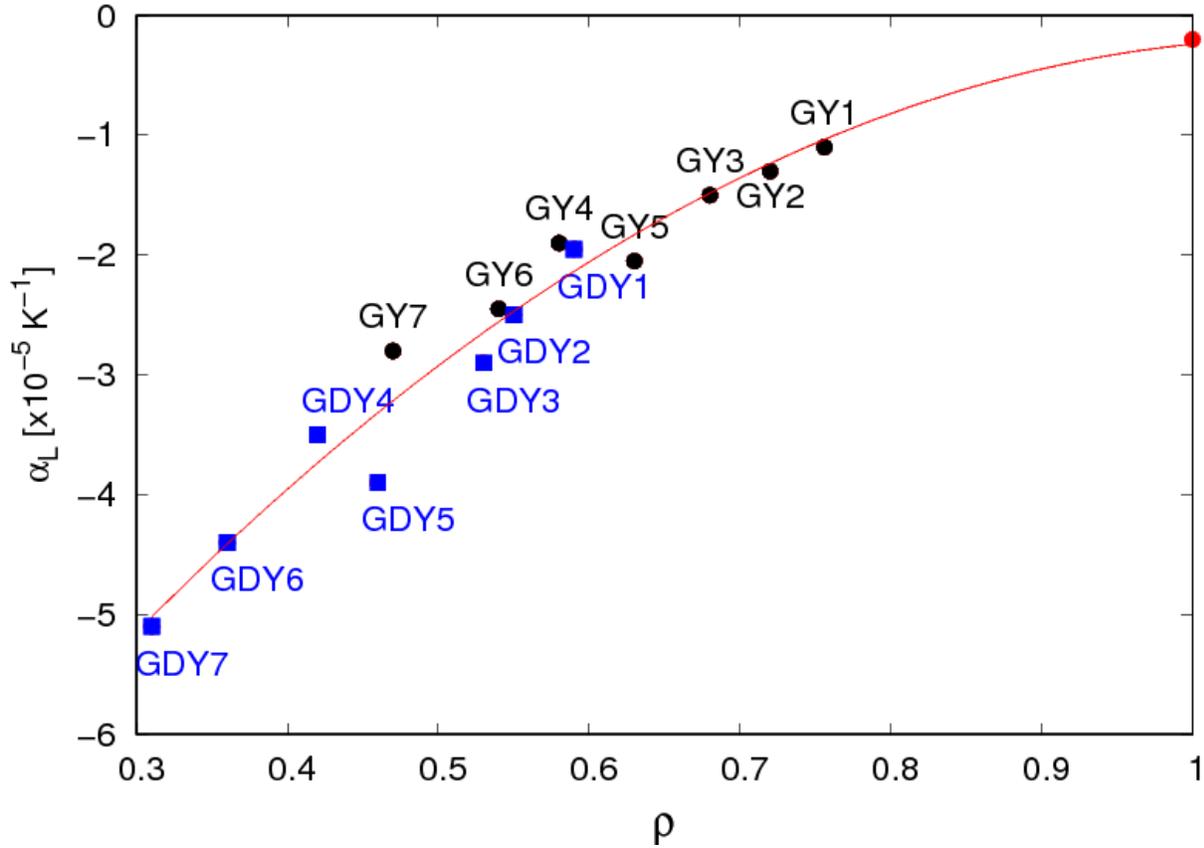

**Figure 3**: (Color online) Average TEC (along *x* and *y* directions) at *T* = 300 K for all GYs and GDYs as function of the relative density, $\rho$. Each point is named for reference and black circles (blue squares) show the data for GYs (GDYs). Red circle at $\rho = 1$ represents the data for graphene. A quadratic fitting of the points is shown by the red full curve.

One can notice that Figures **S3** and **S4** of Supplementary Material [73] show an approximately linear dependence of the equilibrium lengths and the TEC of GYs and GDYs with temperature, different from how the equilibrium length varies with temperature for graphene [66,67]. The linearity in the variation of equilibrium length of the structures with temperature might be also a consequence of the effects of the RUMs, since the increase in temperature leads to an increase only of the amplitude of the RUMs. But, the linear dependence of the TEC with temperature (see Figures **S5** and **S6** of Supplementary Material [73]) is not simply expected from a linear dependence of the equilibrium length $L$ based on the equation (1) (see previous Section). Let us analyze this. If the equilibrium length, $L(T)$, can be written as

$$L(T) = AT + B \ , \tag{7}$$

where $A$ and $B$ are coefficients to be obtained from the fitting of the points shown in Figure **S3** and **S4** of Supplementary Material [73], equation (1) shows that the TEC, $\alpha(T)$, can be simply given by:

$$\alpha(T) = \frac{A}{B}\left(\frac{A}{B}T + 1\right)^{-1} \ . \tag{8}$$

This expression $\alpha(T)$ is not linear in $T$, however if $A/B << 1$, it is possible to show that:

$$\alpha(T) \cong \frac{A}{B}\left(1 - \frac{A}{B}T\right) \ , \tag{9}$$

that is linear in $T$. In our calculations, $A/B \leq \sim 5 \times 10^{-5}$, what justifies the observed linearity of the TEC with temperature with a small inclination, that is what we see in Figures **S5** and **S6** of Supplementary Material [73].

**Young's modulus and Poisson's ratio of GYs and GDYs.** Our results for the Young's modulus and Poisson's ratio of GYs and GDYs are shown in Table **I**. The excellent agreement with the existent data in the literature (data between parenthesis in Table **I**) indicates the correctness of the protocols used here. The available Young's modulus data are for graphene, GY1 (or γ-graphyne), GY2 (or "6,6,12"-graphyne), GY4 (or β-graphyne), GY7 (or α-graphyne), GDY1 (or γ-graphdiyne) and GDY7 (or α-graphdiyne). The curves of $U^{\text{uniaxial}}$ and $U^{\text{biaxial}}$ as functions of tensile strain for all structures, from which we extracted the corresponding values of $C_{11}$, $C_{12}$ and $C_{22}$, are shown in Figures **S7** and **S8** of Supplementary Material [73].

The consistency of our results for the elastic parameters along *x* and *y* directions with the symmetry of the structures are also verified from the data shown in Table **I**. As mentioned in the previous Section, a way to test our protocols and calculations is not previously assuming that the elastic properties of the structures studied here are symmetric. Graphene is known to have hexagonal symmetry and, from Figure **1**, we see that GY1, GY4, GY7, GDY1, GDY4 and GDY7 also have hexagonal symmetry. In order to facilitate the analysis, we call this group of structures as "symmetric group" (SG). The set of non-symmetric structures, GY2, GY3, GY5, GY6, GDY2, GDY3, GDY5 and GDY6, is called "non-symmetric group" (NSG). Table **I** shows that for graphene and every structure of SG, the values of $E_x$ and $E_y$, as well as $v_{xy}$ and $v_{yx}$, are very close one to each other as expected from theory [58]. On the other hand, the results along *x* and *y* directions for the NSG are not close one to each other, thus consistent with the absence of square or hexagonal symmetry. This (ani)isotropy can be seen in the difference between the curves for $C_{11}$ and $C_{22}$ in Figures **S7** and **S8** of Supplementary Material [73]. These results validate our chosen methods.

**TABLE I**. Young's modulus along $x$ and $y$ directions, $E_x$ and $E_y$, respectively, in units of Nm$^{-1}$ and Poisson's ratio $v_{xy}$ and $v_{yx}$, as defined by Eqs. (3) and (4). Data from literature is also included for comparison. Values of Young's modulus or Poisson's ratio from literature without the information regarding the direction were included only in the $E_x$ or $v_{xy}$ columns, respectively.

| Structure | $E_x$ [Nm$^{-1}$] | $E_y$ [Nm$^{-1}$] | $v_{xy}$ | $v_{yx}$ |
|---|---|---|---|---|
| Graphene | 300.28 (242.15 – 424.85)[a] | 287.08 (242.15 – 424.85)[a] | 0.29 (0.14 – 0.40)[a] | 0.28 (0.14 – 0.40)[a] |
| GY1 or γ-graphyne | 163.0 (201.4[b], 150[c], 170.2[d], 237.3[e], 162.1[f], 170.4[g], 166.3[h], 183.8[i], 229.9[j], 165.5[k], 137[l], 166.0[m]) | 159.6 (247.0[b], 169.2[d], 224.0[g], 209.8[j]) | 0.39 (0.26[e], 0.429[f], 0.416[h], 0.396[j], 0.478[l], 0.417[m]) | 0.38 (0.421[j]) |
| GY2 or "6,6,12"-graphyne | 121.1 (117.3[d]) | 152.1 (149.1[d]) | 0.39 | 0.49 |
| GY3 | 98.8 | 141.3 | 0.41 | 0.58 |
| GY4 or β-graphyne | 93.6 (87.1[d], 73.1[k]) | 92.1 (87.4[d]) | 0.52 | 0.51 |
| GY5 | 25.9 | 72.8 | 0.46 | 1.30 |
| GY6 | 32.3 | 55.6 | 0.59 | 0.97 |
| GY7 or α-graphyne | 42.8 (39.9[d], 22.0[k], 22.5[l], 32.0[n]) | 42.4 (40.2[d]) | 0.72 (0.87[k], 0.87[l], 0.72[n]) | 0.72 |
| GDY1 or γ-graphdiyne | 118.6 (100[c], 150.2[g], 123.1[h], 121.8[o], 120.8[p]) | 117.5 (185.2[g]) | 0.40 (0.446[h], 0.453[o], 0.454[p]) | 0.40 |
| GDY2 | 78.5 | 111.7 | 0.39 | 0.55 |
| GDY3 | 59.5 | 102.4 | 0.40 | 0.69 |
| GDY4 | 63.5 | 57.0 | 0.56 | 0.50 |
| GDY5 | 11.1 | 40.2 | 0.44 | 1.60 |
| GDY6 | 14.2 | 22.9 | 0.66 | 1.06 |
| GDY7 or α-graphdiyne | 14.1 (6.7[m], 20.4[q]) | 14.1 | 0.85 (0.91[m], 0.88[q]) | 0.82 |

[a] Refs. [60,78,79], range of values obtained from different methods. [b] Ref. [13]. [c] Ref. [15]. [d] Ref. [16]. [e] Ref. [17]. [f] Ref. [18]. [g] Ref. [19]. [h] Ref. [20]. [i] Ref. [23]. [j] Ref. [24]. [k] Ref. [25]. [l] Ref. [53]. [m] Ref. [69]. [n] Ref. [26]. [o] Ref. [14]. [p] Ref. [58]. [q] Ref. [21].

The literature lacks the values of the Young's modulus of GY3, GY5, GY6, GDY2, GDY3, GDY4, GDY5 and GDY6, and the Poisson's ratio of GY2 (or "6,6,12"-graphyne), GY3, GY4 (or β-graphyne), GY5, GY6, GDY2, GDY3, GDY4, GDY5 and GDY6. A comparison amongst these results shows some interesting similarities and differences that we point out below.

First observation is that the Young's modulus of the structures decreases with the structure density. It is known that the elastic modulus of porous materials grows proportionally to the square of the density [80], $E \sim \rho^2$. As the literature reported that this modulus-density relationship is not valid for some nanostructures [81,82], we investigate the power law dependence of the Young's modulus of GYs and GDYs on their density relative to the density of graphene. Figure **4** show log-log graphics of the variation of the elastic modulus along $x$ (left panel) and $y$ (right panel) directions of GYs and GDYs with the relative density of the structure. By fitting the points by the expression $E_x \sim \rho^{px}$ and $E_y \sim \rho^{py}$, the exponents $px$ and $py$ for all, SG and NSG structures were obtained and shown in Table **II**.

Table **II** shows the exponents of power law relationships between Young's modulus and density of all, only the SG and only the NSG ones. In all cases, the exponent of modulus-density relation is larger than 2, so indicating larger sensibility of the Young's modulus of GYs and GDYs with density than that of regular porous materials [80]. This result might be related to the fact that in graphene, the carbon-carbon bonds are all $sp^2$ – hybridized while in GYs and GDYs the acetylenic and diacetylenic linkages provides different elastic behavior to the structures. Symmetric structures present a symmetric value for the exponent as expected. But the large asymmetry between $px$ and $py$ observed within non-symmetric structures (NSG) are remarkable.

This result correlates to the observation that for all NSG structures, the in-plane stiffness along *x* (armchair) direction are always smaller than that along *y* (zigzag) direction (see Table **I**). So, in NSG GY and GDY structures, the armchair direction not only is the softest one, but also presents an exponent of modulus-density relation resembling that of highly porous gels materials [81,83].

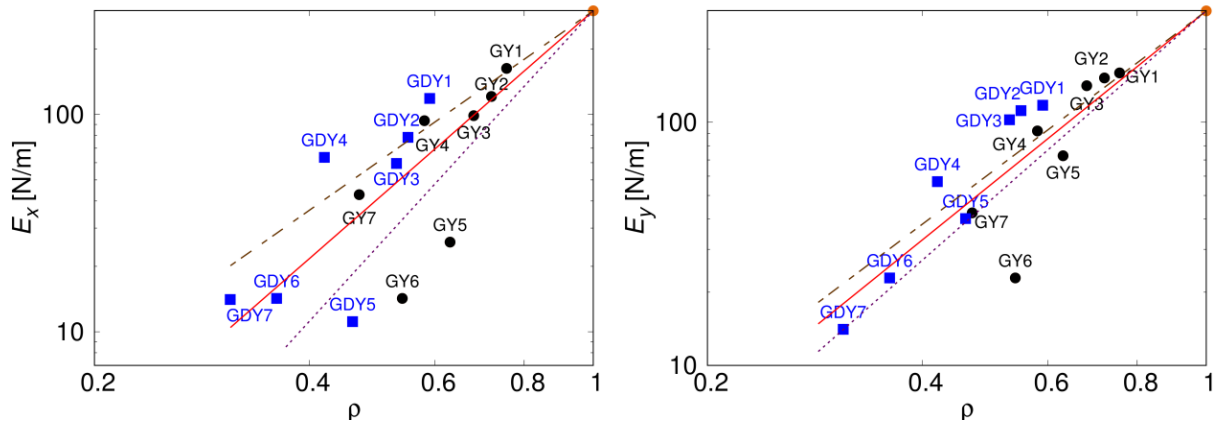

**Figure 4**: (Color online) Young's modulus along *x* (left panel) and *y* (right panel) directions of all GY and GDY structures as function of the relative density, $\rho$. Black circles and blue squares represents the data corresponding to GYs and GDYs, respectively. Red circle on top right side of the graphics represents the values for graphene. A linear fitting of all data, data for symmetric structures (SG) and data for non-symmetric structures (NSG) are drawn in red full, black dashed and black dotted lines, respectively. The angular coefficients of all lines corresponding to the exponents of the relation $E \sim \rho^p$ are given in Table **II**.

**TABLE II**. Exponents of modulus-density relations $E_x \sim \rho^{px}$ and $E_y \sim \rho^{py}$, for all, the symmetric (SG) and non-symmetric (NSG) GYs and GDYs.

|     | *px* | *py* |
| --- | --- | --- |
| ALL | 2.87 | 2.37 |
| SG  | 2.31 | 2.20 |
| NSG | 3.59 | 2.58 |

Regarding the Poisson's ratio of GYs and GDYs, the first observation is that all of them are larger than that of graphene. This indicates that they are softer than graphene. Amongst the symmetric structures (structures from SG), the Poisson's ratio varies from about 0.39 to 0.85. This is compatible with the prediction from equation (6) that $v_{xy}$ cannot be larger than 1. These values are also compatible with those predicted for transversely isotropic materials for which $v_{xy}$ can vary from -1 to 1 [84]. Some of GY and GDY structures present Poisson's ratio very close to 0.5, (GY1, GY4, GDY1 and GDY4). Poisson's ratio equals 0.5 is a characteristic of an incompressible material (like rubber) which conserve volume when subjected to axial strain. This feature was observed by Peng *et al*. [18] for GY1 (or γ-graphyne) but that is found here to be also true for GY4, GDY1 and GDY4 structures.

Amongst the symmetric structures, GY7 (or α-graphyne) and GDY7 (or α-graphdiyne) are the ones with the largest values of Poisson's ratio, ~ 0.72 and 0.85, respectively. This result might be interpreted as follows. According to Christensen [84], incompressible materials which possess an infinitely rigid stiffness along one particular direction, present Poisson's ratio equals 1 along the two other directions orthogonal to the first one. The values of Poisson's ratio of GY7 and GDY7, 0.72 and 0.85, respectively, suggests that these structures behave similarly to incompressible materials with rigid stiffness along the direction orthogonal to their plane.

For the structures in NSG, not only the Poisson's ratio along the two planar directions are asymmetric (as expected), but also they can reach values larger than 1. Although these structures are complex, the simple honeycomb model depicted by figure **2** can help understand the underlying physics of, at least, one of the non-symmetric cases.

A simple model of a planar honeycomb formed by rigid ribs that can stretch and/or hinge, predicts $v_{xy} = v_{yx}$ and the upper limit of 1 for the Poisson's ratio of symmetric structures (see equation (6) in previous Section). However, the model depicted in figure **2** cannot be directly applied to GY2, GDY2, GY3, GDY3, GY6 and GDY6 NSG structures. Fortunately, we can identify the model with the structure of GY5 and GDY5, as long as we consider different values of the rib lengths *l* and *h*. An acetylenic or diacetylenic chain (–C≡C– or –C≡C–C≡C–, respectively) has three or five carbon-carbon bonds, respectively. Let's consider the approximate ratios of *l*/*h* = 3 and *l*/*h* = 5 for GY5 and GDY5, respectively, where *l* is the length of the corresponding acetylenic or diacetylenic chain, and *h* is the length of the carbon-carbon double bond parallel to *x* direction (see figures **1** and **2** to identify the GY5 and GDY5 structures with the honeycomb model). Then, let us consider that the elastic constant due to stretching of the ribs, $K_S$, in the honeycomb model, is about 10 times larger than the elastic constant due to hinging of adjacent ribs, $K_H$, i. e. $K_S \sim 10 K_H$. This estimate is at the same order of magnitude of what can be inferred from numbers and equations for constant force parameters related to stretching and bending of carbon-carbon bonds in graphene [85]. Then, let us substitute the above values for *l*/*h* and $K_S/K_H$ on the equations (5), for GY5 and GDY5. The results are $v_{xy}^{GY5} \cong 0.45$, $v_{yx}^{GY5} \cong 1.25$, $v_{xy}^{GDY5} \cong 0.43$ and $v_{yx}^{GDY5} \cong 1.77$. These values are in excellent agreement with those obtained from the MD simulations (from Table **I**: $v_{xy}^{MD-GY5} = 0.46$, $v_{yx}^{MD-GY5} = 1.30$, $v_{xy}^{MD-GDY5} = 0.44$ and $v_{yx}^{MD-GDY5} = 1.60$). This suggests that the elasticity of GY5 and GDY5 structures are mainly resulting of stretch and hinge degrees of freedom, at least, when subjected to tensile strains.

## IV. CONCLUSIONS

We have performed a comparative study of the thermal expansion and elastic properties of all seven originally proposed [3] GYs and the corresponding GDYs by means of classical MD simulations. The TEC at 300 K of all GY and GDY structures was shown to be, in modulus, one order of magnitude larger than that of graphene and the thermal contraction quadratically increases with structure density.

The Young's modulus of the GY and GDYs structures are all smaller than that of graphene and shown to increase with the structure density through a power law relationship with exponent larger than that of regular porous materials. An interesting result was observed for non-symmetric GY and GDY structures. It was not only shown that these structures present anisotropic elastic behavior, but also that the elastic modulus along their armchair direction is always smaller than that along zigzag direction. The modulus-density relationship along the armchair direction of these non-symmetric structures were shown to be much more like porous gels materials than regular porous materials. It indicates that for such structures the mechanical properties degrade much faster with increasing porosity than that of symmetric group. It will be interesting to verify how this trend changes by considering graph*tri*yne and graph"*more*"yne structures.

The Poisson's ratio of GY and GDY structures are all larger than that of graphene, what indicates that these structures are softer than graphene. The values of the Poisson's ratio of GY and GDY structures along armchair and zigzag directions are consistent with their symmetry, and a honeycomb model formed by ribs that are able to stretch and hinge is capable of predicting the results. The honeycomb model with two different sizes of ribs resembling the non-symmetric

GY5 and GDY5 structures, was shown to quantitatively predict the anisotropic values of their Poisson's ratios. It is remarkable that a simple honeycomb ribs-based model can capture the elastic behavior of molecular planar structures as GYs and GDYs.

## SUPPLEMENTARY MATERIAL

See supplementary material for Figures **S1** - **S8** regarding the tests of convergence of calculation of TEC of GYs and GDYs with size; values of the equilibrium length and the TEC of the structures as functions of temperature; and the curves of the variation of the energy of the structures with uniaxial and biaxial tensile strains.

## ACKNOWLEDGMENTS

S. A. H. was supported by the Brazilian agency CAPES. A.F.F. is a fellow of the Brazilian Agency CNPq (#302750/2015-0) and acknowledges grant #2016/00026-9 from São Paulo Research Foundation (FAPESP). This research used the computing resources and assistance of the John David Rogers Computing Center (CCJDR) in the Institute of Physics "Gleb Wataghin", University of Campinas.